\documentclass[12pt]{elsarticle}
\usepackage{graphicx} 
\usepackage{amsmath,amsfonts,amssymb,bm}
\newcommand{\Z}{\mathbb Z}
\newcommand{\appreq}{\simeq}

\newcommand{\submod}[1]{\mathbin{\ominus_{#1}}}
\newcommand{\Eq}[1]{Eq.~(\ref{#1})}
\newcommand{\Fig}[1]{Figure~\ref{fig:#1}}
\newcommand{\Figs}[3]{Figures~\ref{fig:#1}#2~\ref{fig:#3}}
\newcommand{\Sec}[1]{Sec.~\ref{sec:#1}}
\newcommand{\CET}[2]{C_{#1}E_{#2}}
\newcommand{\WL}[1]{\CET0{#1}^{\,I\!I}}
\newcommand{\vNT}[1]{E_{#1}}
\newcommand{\REd}{\vNT{1,3}^{I\!I}}
\newcommand{\wt}[1]{wt({#1})}
\newcommand{\Dfr}{{\mathfrak D}}
\newcommand{\Ddfr}{\Dfr^\diamond}

\journal{Physica D (Nonlinear Phenomena)}

\begin{document}
\begin{frontmatter}

\title{Modelling reliability of reversible circuits with
2D second-order cellular automata}
\date{\today}
\author{Alexander Yu.\ Vlasov}

\begin{abstract}
The cellular automaton is a widely known model of both reversible 
and irreversible computations.
The family of reversible second-order cellular automata 
considered in this work is appropriate
both for construction of logic gates and analysis of damage 
distribution. The quantities
such as formal dimension of damage patterns 
can be used only for rough estimation of consequences of
particular faults and numerical experiments are provided for illustration 
of some subtleties. Such analysis demonstrates 
high sensitivity to errors from defects, lack of synchronization
and too short intervals between signals.
\end{abstract}

\begin{graphicalabstract}
\includegraphics[width=\textwidth]{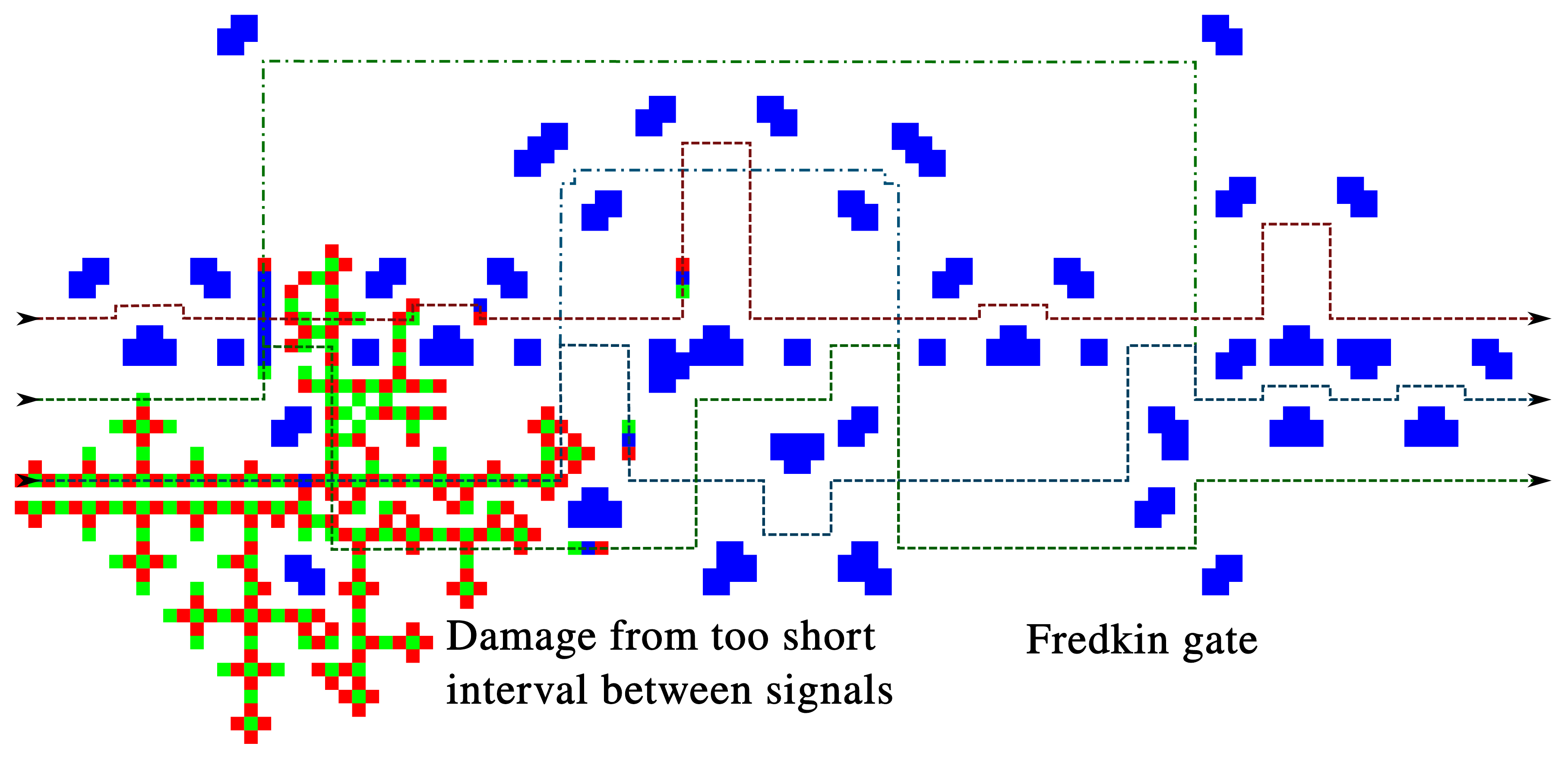}
\end{graphicalabstract} 



\end{frontmatter}

\section{Introduction} 
\label{sec:Intro}

The reversible computations have quite a long history \cite{Ben73} and nowadays 
they deserve an increasing attention due to different applications 
from the reduction of power consumption to the theory of quantum 
computations \cite{KM17,COST20,RC22,RC23}.
The analysis of uncertainties for the testing of usual logic circuits  
is well developed area with variety of different methods \cite{Circ13}, 
but some subtleties can appear for the reversible circuits.
The problems of faults and the reliability of the reversible circuits 
were also covered quite widely \cite{PHM03,HPB04,RTest20}.

Sometimes the testing of reversible circuits may be even simpler for
some particular faults \cite{PHM03}, but some difficulties may appear in the more
general cases \cite{HPB04}. There are also some problems due to specific properties
of the reversible circuits. From the one hand, the reversibility produces an advantage
due to negligible power consumption, from the other one, the undesirable 
perturbations may spread without dissipation and produce the problems with too high 
error sensitivity.

More detailed analysis of the particular structures used for construction 
of the reversible logic circuits can be useful for the sensitivity analysis 
and consideration of some other problems. A model of so-called quantum-dot
cellular automata can be used as some example \cite{RTest20,QDCA07}.
The presented paper is devoted to the more abstract models based on 
the reversible cellular automata (RCA).

The consideration of the uncertainty propagation can be described with
irreversible probabilistic cellular automata \cite {UQCA}, however
an application of similar ideas to the reversible case produces some problems
if to try add the noise directly into update rule of the cells, because such a 
system could not perform the reliable computations \cite{RCAnoise}.
 
The issue with noisy local rules of the probabilistic RCA justifies 
sensitivity analysis of usual (deterministic) RCA evolution with 
respect to different perturbations of the initial configurations. The family of
second-order RCA with four states considered in this work uses the standard 
method of construction from the cellular automata with 
two states \cite{M84,V84,TM90}.

The presented work is in some agreement with an earlier suggestion from 
Ref.~\cite{TM90} `[\ldots] the most practical approach is to start with 
a local map that directly supports logic gates and wires, and then build 
the appropriate logic circuits out of these primitives'. 
Examples of such approach often inspired by so-called 
{\em billiard ball model} (BBM) \cite{KM17,FT82,JK18}. A computationally 
universal second-order RCA derived from cellular automaton with three 
states was discussed already in Ref.~\cite{M84}.

Another approach to the computational universality in RCA uses
1D design based on universal Turing machine and it is discussed together 
with 2D logic circuits based on so-called partitioned CA 
(PCA) in Ref.~\cite{KM17,JK18}. PCA is usual choice for modelling
of reversible circuits due to natural relation with BBM. 

The conservation of number of occupied cells can be simply implemented
in PCA directly by local transition rules preventing some problems 
with damage distribution discussed further. On the other hand, more 
realistic models of logic circuits may require possibility
to investigate noise and damage.

Models without conservation of cell number may be necessary in such a case.
In irreversible CA undesirable rapid growth of number of cells could be 
corrected by modification of local rules to cancel some configurations.
For example, diverse evolution {\em Game of Life} CA  \cite{WW,Gard}
and possibility of universal computations with such a model is supported by
rules with good balance between appearance and cancellation of 
different configurations. 

However, there is specific problem with RCA, because reversible evolution
makes impossible extinction of any configuration and without conservation 
of cell number (or some other non-negative quantity) undesirable growth and 
chaotic behaviour of initially regular configuration is rather common 
result of RCA evolution in unbounded domains. 

A few reversible second-order RCAs discovered by author some time ago and
discussed in this work provide interesting combination of
stable and expansive behaviour. From the one hand such RCA can be used for
creation of reversible universal gates, from the other one even expansive 
behaviour displays some regularities and fractal patterns.
Investigation sensitivity and reliability of reversible circuits with
such models can be quite instructive and relevant to analogue
processes in more realistic systems.

\smallskip

Let us briefly outline contents of this paper.
After recollection of basic properties and possible classification
of CA in \Sec{CAbas} and \Sec{class} a few original
second-order RCAs for circuits construction are introduced in
\Sec{CAC0} and examples of logic gates for universal computations
are described in \Sec{gates}.
Different kinds of errors from the defects and
the improper signals interactions are discussed in \Sec{damage}  
recollecting some quantities such as {\em Lyapunov exponents} 
and {\em formal dimensions} those could be useful for 
analysis of the damage spread together with the numerical experiments. 
In conclusion, \Sec{fin} summarizes this work.

\section{Reversible cellular automata}
\label{sec:CA}

\subsection{Basic properties and classification of CA}
\label{sec:CAbas}

Let us recollect some properties of cellular automata (CA) on the rectangular
grids with dimension $d$ often denoted as $\Z^d$ \cite{WolCA,JK05}. Two-dimensional
CA are quite natural for modelling of the logic circuits. The {\em state} of any
cell on the grid $\Z^d$ belongs to the finite-dimensional set $S$ with $k$ elements
and for simplicity may be described by a number from interval $0,\ldots,k-1$.
State $k=0$ could be considered as an empty cell and there are $k-1$ kinds
of occupied cells. In the simplest case with two states the cell is either 
occupied or not.

Evolution of such a system is described by synchronous update at discrete
time steps. The {\em local update rule} for each cell $c_{\mathbf z}$,  
$\mathbf z \in \Z^d$ is a function dependent on only a few cells in the certain 
neighborhood described by $n$ vectors of relative displacements
$\mathcal N = (\mathbf v_1,\ldots,\mathbf v_n)$
\begin{equation}
\label{fCA}
c_{\mathbf z}^{(t+1)} = f(c_{\mathbf{z+v}_1}^{(t)},\ldots,c_{\mathbf{z+v}_n}^{(t)})
\doteq f(c_{\mathbf z+ \mathcal N}^{(t)}).
\end{equation}
There are two basic kinds of neighborhoods for 2D rectangular grid. The cell itself 
with four closest cells with common edge form the {\em von Neumann} neighborhood  $(n=5)$ 
and the {\em Moore} neighborhood includes additional four cells with common corner $(n=9)$.

The reversible cellular automata (RCA) are appropriate for the purposes of presented
work, but there are no effective algorithms
for inverting given local rule for $d>1$ \cite{JK18}. However, some methods
allow to construct new local rules with known simple inverse from the very beginning. 

One of these methods uses so-called second-order CA \cite{M84,V84,TM90,JK18,Ila} 
recollected below. The second-order CA uses values from two previous steps for
calculating of a state for a next one
\begin{equation}
\label{f2ndCA}
c_{\mathbf z}^{(t+1)} = f_2(c_{\mathbf z+ \mathcal N}^{(t)},c_{\mathbf z+ \mathcal N}^{(t-1)}).
\end{equation}

Let us now consider arbitrary CA with a local update rule $f$ \Eq{fCA}. It is possible
to define new second-order CA 
\begin{equation}
\label{f2RCA}
c_{\mathbf z}^{(t+1)} = f(c_{\mathbf z+ \mathcal N}^{(t)}) \submod{k} c_{\mathbf z}^{(t-1)},
\end{equation} 
where $\submod{k}$ is subtraction modulo $k$ and such CA is reversible \cite{M84,V84,TM90,Ila}
with the inverse rule
\begin{equation}
\label{Inv2RCA}
c_{\mathbf z}^{(t-1)} = f(c_{\mathbf z+ \mathcal N}^{(t)}) \submod{k} c_{\mathbf z}^{(t+1)}.
\end{equation} 

For an initial CA with two states $k=2$ the operation $\submod2$ coincides with XOR
(eXclusive OR often denoted as $\oplus$)
producing more familiar examples \cite{JK18}. The second-order RCA can be considered
as first-order RCA with state spaces $S \times S$ and local rule \cite{TM90,JK18}
\begin{equation}
\label{SqRCA}
\bigl(c_{\mathbf z}^{(t-1)},c_{\mathbf z}^{(t)}\bigr) \mapsto
\bigl(c_{\mathbf z}^{(t)},c_{\mathbf z}^{(t+1)}\bigr) =
\bigl(c_{\mathbf z}^{(t)},f(c_{\mathbf z+ \mathcal N}^{(t)}) \submod{k} c_{\mathbf z}^{(t-1)}\bigr).
\end{equation}

For CA with $k$ states such a method produces RCA with $k^2$ states.
Here is considered simplest case of CA with two states $k=2$ and 
$k^2 = 4$ states for RCA. The states can be considered either
as pair of bits $(s^{(1)},s^{(2)})$, or as single number $s= s^{(1)}+2s^{(2)}$,
$0 \le s \le 3$. Thus, for such RCA non-empty cells can have three different states
$s=1, 2, 3$, sometimes denoted further for visibility as red, green and blue
respectively. 

\subsection{Classification of CA and RCA}
\label{sec:class}

Due to \Eq{SqRCA} there is unique relation between the second-order RCA 
with four states and two-state CA. Thus, classification of 2D CA with 
two states can be also used for such RCA. 
A natural choice is the {\em isotropic}
local rules also known as {\em completely symmetric} \cite{PW85}, {\em i.e.,}
invariant with respect to rotations and reflections. However, even with
such restriction there are $2^{102} \appreq 5 \times 10^{30}$ different CA \cite{PW85}.

There are more narrow classes of CA, for example {\em totalistic} with
value $f$ in \Eq{fCA} is depending only on sum of occupied cells in neighborhood and
{\em outer totalistic} with taking into account the sum and the 
cell itself \cite{PW85}. The last example is also known as
{\em semi-totalistic} or {\em Life-like} CA \cite[Ch.~6]{GoL10} due to 
famous Conway's {\em Game of Life}\/ CA \cite{WW,Gard} and there
are $2^{18} \appreq 3 \times 10^5$ such rules \cite{PW85}.

For the certainty the term {\em outer totalistic} is saved further for 
particular case of {\em outer totalistic inner-independent} CA \cite[Ch.~13]{GoL10} 
with local rule taking into account only `outer neighborhood' {\em without cell itself}.

Between isotropic and totalistic classes there is an intermediate one with separate counts 
for two kinds of neighboring cells: the four closest neighbors have a common edge with 
central cell and another four cells have only common corner. Such a class is also known as 
{\em corner-edge totalistic} (CET) or {\em quarter-totalistic} CA rules \cite{term} and 
it is convenient again to consider {\em inner-independent} case without dependence 
on central cell denoted further as {\em corner-edge outer totalistic} (CEOT) CA.

There are $2^{50} \appreq 10^{15}$ CET CA and $2^{25} \appreq 3 \times 10^7$ CEOT CA.
Indeed, amounts of non-empty cells with common corners and edges is pair of numbers
$(n_c,n_e)$, where $0 \le n_c,n_e \le 4$ and there are 25 possible pairs.
The local rule $f$ for CEOT can be described by a sequences $s$ of such pairs $p$ 
with property $f(p)=1$ {\em if and only if} $p \in s$. There are $2^{25}$ such sequences
and different CEOT rules. There are $(2^{25})^2 = 2^{50}$ different CET rules taking into 
account two possible states of central cell.

The similar method can be used to calculate number of possible local rules in other
classes of CA with two states discussed above and corresponding second-order RCA.
The considered classes also can be convenient due to relations
with invariant quantities in second-order RCA defined by
count of cell values \cite{V84,Ila,YP84,GV86,ST12}. 
However, there are few alternative methods of classifications \cite{class} 
and these topics should be discussed elsewhere. 

\subsection{RCA family for circuits construction}
\label{sec:CAC0}

Let us consider subfamily of CEOT CA with local update rules producing occupied cell only if
all cells with common corner are empty, {\em i.e.}, only pairs with $n_c=0$
may appear in the sequence for $f(p)=1$ discussed above. 
Selected CA are between simplest examples of CA from considered families (CET, CEOT) and 
were found by not very extensive search for CA with universal gates. Anyway, analogue 
CA were not found in similar way between totalistic rules and so CEOT was used instead.

The simple example of such CA denoted here as $\CET01$ has local rule with $f(p)=1$ only 
for single pair $(n_c=0, n_e=1)$ and second-order RCA derived
from such a rule using \Eq{SqRCA} is denoted as $\WL1$. 
Few different extensions $\WL{1\dots}$ of such RCA defined further include 
more pairs with $n_c=0$, $n_e \ge 1$ in the sequence for $f(p)=1$.

Condition about empty cells with common corner is quite restrictive
and many configurations of initial irreversible CA disappear after single step,
but in \Eq{SqRCA} for second-order automata $f(p)=0$ is applied
as XOR operation to the previous state.
Configurations corresponding to 
those that would disappear in the original CA with two states become
stable in the second-order RCA if all cells have the state $s=3$ (with two 
units in binary notation) or blink between two values for $s=1,2$.

The RCA $\WL1$ also has moving configurations appropriate for encoding of signals 
in the circuits. There are infinite amount of possible signals represented
as lines with length $k \ge 3$ starting with red cell (state $s=1$) and ending 
with a green cell (state $s=2$) with $k-2$ blue cells (state $s=3$) in between
(`snake').
If to use second-order description with two layers such configurations
correspond to two lines with length $k-1$ with one cell displacement  
in direction of motion. The speed of such signals is one cell in vertical or horizontal direction per time step.

The signal can change direction (and length) after interaction with stable 
configurations. Signals with odd length can change direction (and length)
after collision with static configurations of blue cells and appear
more convenient for control than signals with even length that needs
for blinking configurations and time synchronization for such control.

\subsection{Constructions of reversible gates}
\label{sec:gates}

An analogy of BBM computations \cite{FT82} can be developed
with shortest three-cell signals (`ants'). The simplest encoding
is supposed in such a case with the unit and zero are represented by presence or absence 
of an elementary signal respectively. It has some disadvantages, but demonstrates
many basic principles of circuits design and more difficult encodings should be 
discussed elsewhere. 

For example, two kinds of signals could be used for unit and zero. 
However, it would require CA with bigger number of states and an alternative 
approach to construction of the gates. Yet another method is to use
signals with different length for unit and zero, but some RCA and gates 
discussed below may work only with signal of minimal length. For so-called dual-rail encoding 
representing both unit and zero as an elementary signal travelling along two possible 
parallel paths the same RCA could be used, but constructions of gates  
would be different.

\smallskip

The important element for BBM is the reversible {\em Fredkin} gate \cite{FT82}. Such a gate (with
three inputs and three outputs) performs conditional exchange between two {\em data} 
signals depending on value of a {\em control} one. 
The action of Fredkin gate can be formally expressed as
\begin{equation}
\label{fred}
 F\!_R : \{A,B,C\} \to \{A,(\lnot A\land B) \lor (A \land C), (A\land B) \lor (\lnot A \land C)\}.
\end{equation}
Here two data signals are exchanged if control signal presents. Such convention
formally differs from initially suggested in Ref.~\cite{FT82}, but it can be simply
altered by exchange of data outputs.

\begin{figure}[htb]
\begin{center}
 \parbox[c]{0.32\textwidth}{
 \includegraphics[width=0.3\textwidth]{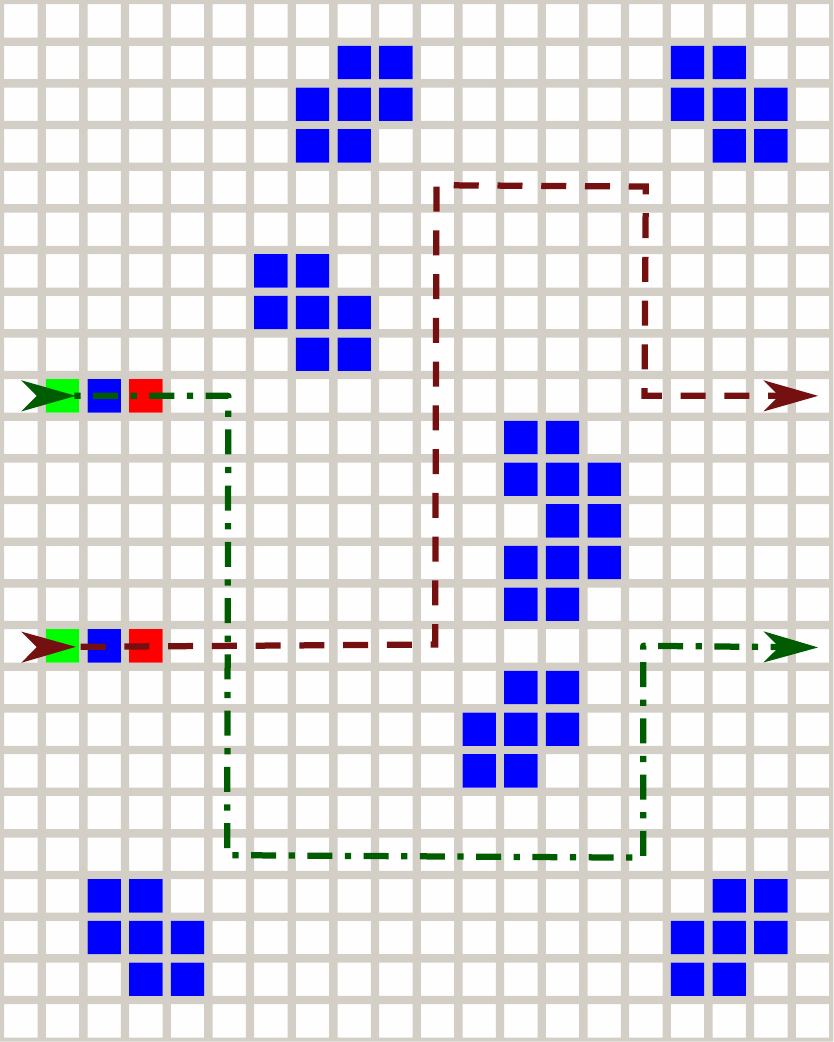}}
 \parbox[c]{0.67\textwidth}{
 \includegraphics[width=0.65\textwidth]{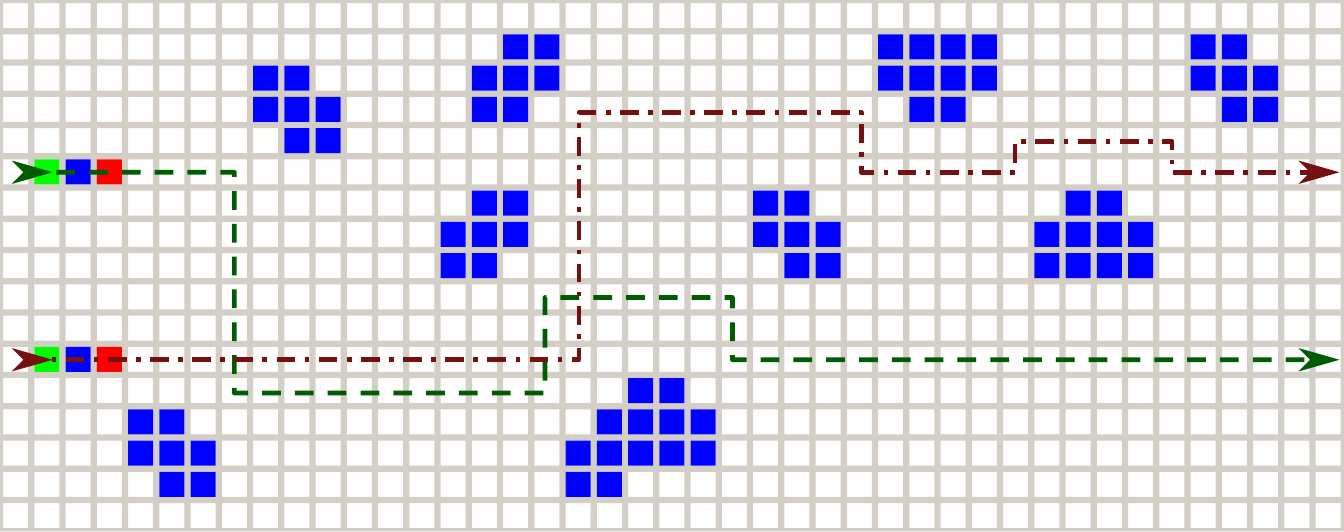}}
\end{center}
\caption{Swap gates for $\WL{1\dots}$: `wide' (left) and `narrow' (right) }
\label{fig:swap}
\end{figure}

The exchange (swap) gate is a simplest element, but for 2D CA some efforts
are required to prevent collision of signals during exchange.
Two possible designs with `wide' and `narrow' swap gates are shown on 
\Fig{swap}. 

Despite of relative simplicity of `wide' swap, 
the `narrow' swap may be reasonable for composition of many
gates with distances between parallel signal paths equal 
to some fixed value, {\em e.g.} $\delta = 6$. 

The pictures \Fig{swap} also illustrate specific way of interaction between signals 
and static elements: there is always an one-cell gap between them. The signal stops, 
because new cell may not appear near static elements and starts motion
in an orthogonal direction. To prevent unlimited growth one of 
the directions should be limited by a design of the static elements.

The Fredkin gates together with delay elements, swaps and `constants' ({\em i.e.} 
some amounts of auxiliary signals) allow to construct any logic circuits \cite{FT82}.
Necessity of some constants for used encoding is clear even from construction 
of NOT gate that can be expressed with Fredkin gate \Eq{fred} as 
$$F\!_R : \{A,1,0\} \to \{A,\lnot A,A\}.$$

It should be mentioned that changing direction on $90^\circ$ 
after reflection in considered RCA occupies some time (unlike BBM). For shortest signal the additional 
delay is one time step and points of grid critical for synchronization of arrival times 
should be chosen accordingly. 

\begin{figure}[htb]
\begin{center}	
  \includegraphics[scale=0.75]{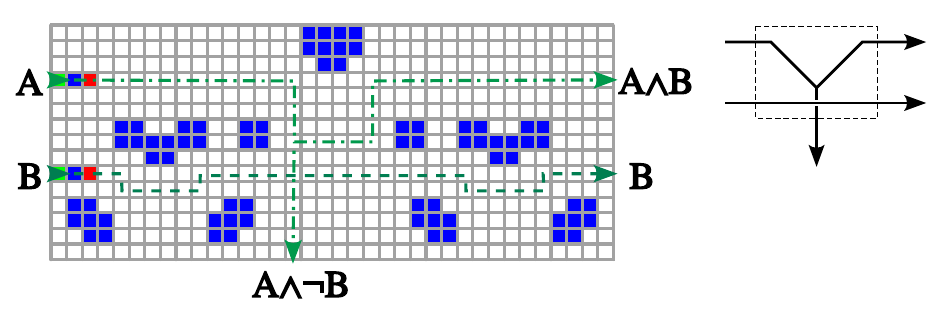}
\end{center}
\caption{Signals routing and collisions in a model of switch gate for $\WL{1\dots}$}
\label{fig:switch}
\end{figure}

Similarly with BBM construction of the Fredkin gate can be expressed with simpler
{\em switch gates} \cite{FT82} for conditional routing of one data signal by a control one.
The configuration of the switch gate for discussed RCA is shown on the left side of \Fig{switch}. 
The scheme of the switch gate shown on the right side of the figure has been modified compared 
to that used in \cite{FT82} to more accurately display the directions of outgoing signals.

Yet another property of considered RCA is used here: after interaction
between two signals with orthogonal direction and proper time delay
one of them continue motion without change, but other one changes
direction on $90^\circ$. The scheme also includes static elements necessary
for direction change of data signal and delay elements to keep synchronization 
of two signals.

\begin{figure}[htb]
\begin{center}	
  \includegraphics[scale=0.2]{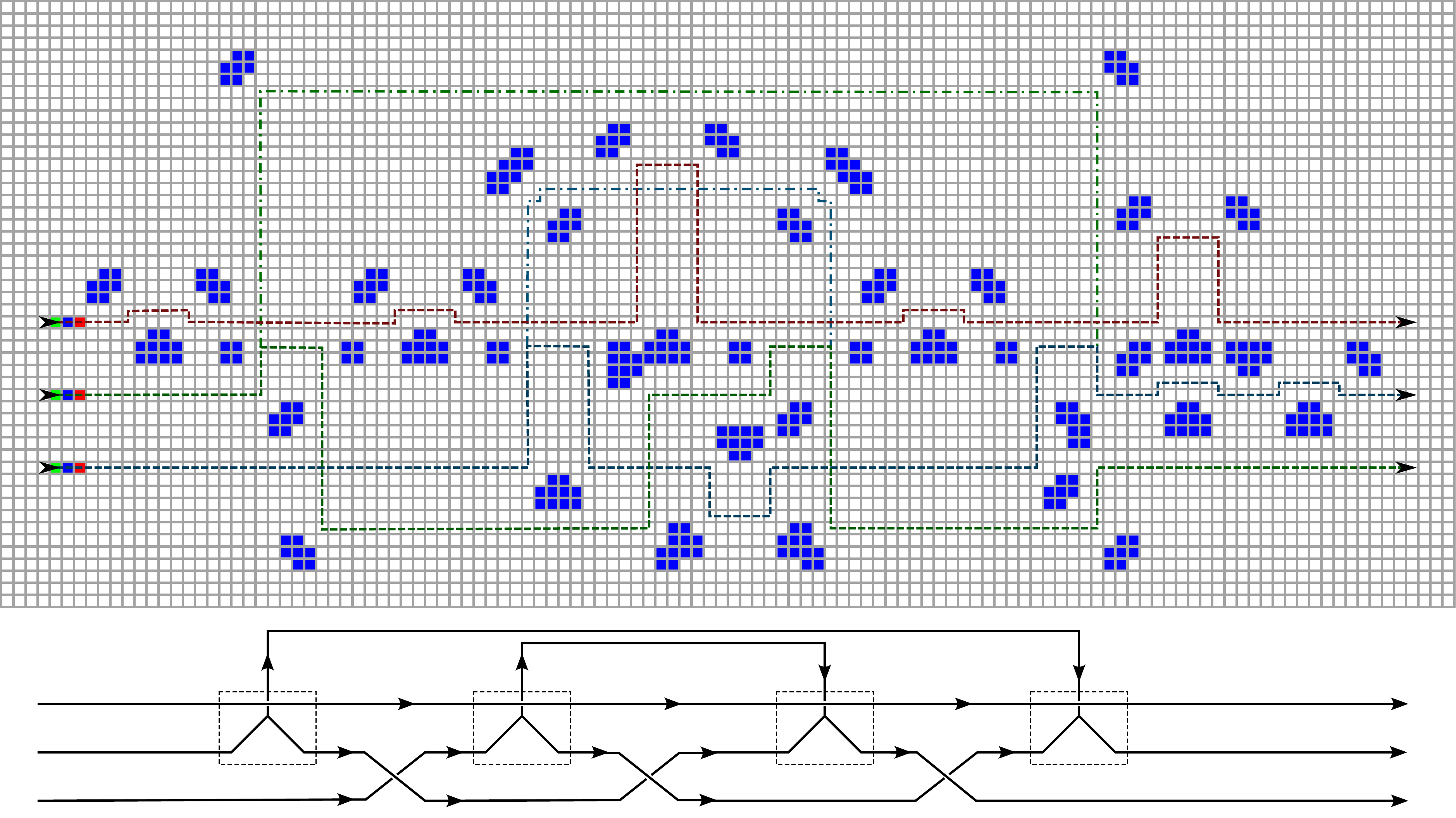}
\end{center}
\caption{Signals routing and collisions in a model of Fredkin gate for $\WL{1\dots}$}
\label{fig:fredkin}
\end{figure} 

The scheme of Fredkin gate composed from four such switch gates initially
proposed by R. Feynman and A. Ressler may be found in Ref.~\cite{FT82}.
Possible realization for considered RCA is shown on \Fig{fredkin}.
The simplified circuit diagram without delay elements shown at 
the bottom of \Fig{fredkin} uses two switch gates together with 
two inverted switch gates. 
All switch gates on \Fig{fredkin} are flipped vertically compared with 
\Fig{switch} and surrounded by dashed rectangles on the diagram.

Without control signal two data signals follow pair of bypass paths 
(above the main diagram) and simply leave the scheme with some delay, but due to
interaction with control signal data signals follow other (triple-crossing) paths 
to exchange output positions. Many static element here is for synchronization
of delays for arbitrary combinations of input signals.

Similarly with BBM the number of input and output signals for RCA $\WL1$ should
be equal. However, minimal modification of local rule allows to avoid such constraint.
Let us save requirement about absence of occupied cells with common corner
in description of local update rule of initial CEOT CA, but number of 
occupied cells with common edge now may be one or two. 
Second-order RCA produced using \Eq{SqRCA} from such
rule is denoted here $\WL{1,2}$.

Signals in modified RCA may have only minimal length (three cells). However,
such kind of signals was also used in previous RCA and there is no difference with
discussed examples such as switch and Fredkin gates. An essential property
of new RCA is possibility to split signals. It was already mentioned, that
after encountering the static obstacles signal start motion in orthogonal
direction, but if both such directions are not limited by some elements,
new RCA instead of stretching produces two signals moving
in opposite sides.

\begin{figure}[htb]
\begin{center}	
  \includegraphics[scale=0.72]{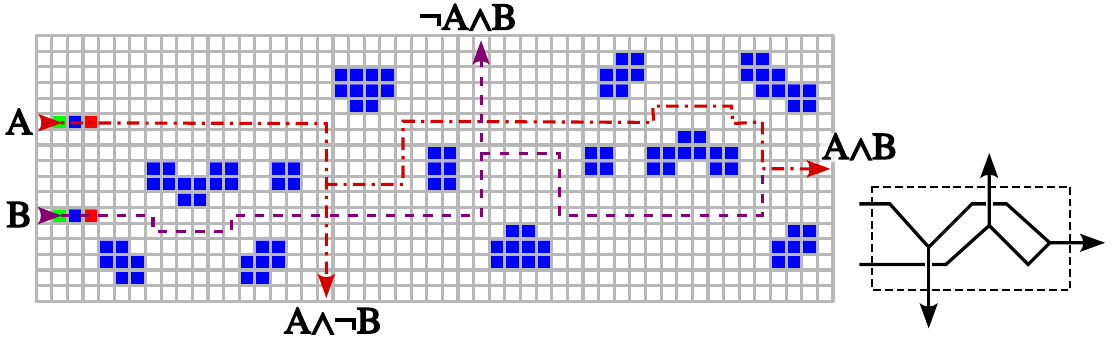}
\end{center}
\caption{Transformation of two signals into three paths in RCA $\WL{1,2}$ }
\label{fig:m-switch}
\end{figure}

Such property provides a counterexample to some confusion about impossibility to 
split signals in reversible system. The inversion of splitting element is merging
of two signals. Let us consider yet another element useful for construction
of Controlled-NOT gate. Similarly with switch gate it has two inputs and 
three output paths, but it may merge two of signals into one path. 
It encodes three possible non-empty combinations of two signals
into one signal moving along one of three possible paths, see \Fig{m-switch}.
The simplified scheme of such gate is shown on the right side of the figure.

Controlled-NOT gate applies NOT to data signal if control signal presents
\begin{equation}
\label{c-not}
 \{A,B\} \to \{A, A \oplus B\},
\end{equation}
where $\oplus$ is addition modulo 2 (XOR).
Such a gate can be represented as
\[
\begin{array}{c|c}
\text{control} & \text{data} \\ \hline
0 & 0 \\
0 & 1 \\
1 & 0 \\
1 & 1 \\
\end{array}
\to
\begin{array}{c|c}
\text{control} & \text{data} \\ \hline
0 & 0 \\
0 & 1 \\
1 & 1 \\
1 & 0 \\
\end{array}
\]

If to consider Controlled-NOT gate as a function on pairs of bits, 
it exchanges rows $(1,0)$ and $(1,1)$ in the table above. 
It can be implemented
by encoding pairs of signals into three different paths using scheme on \Fig{m-switch},
exchanging two appropriate paths and applying the inverse of the scheme
(with split of signal along one of paths). 

\begin{figure}[htb]
\begin{center}	
  \includegraphics[scale=0.2]{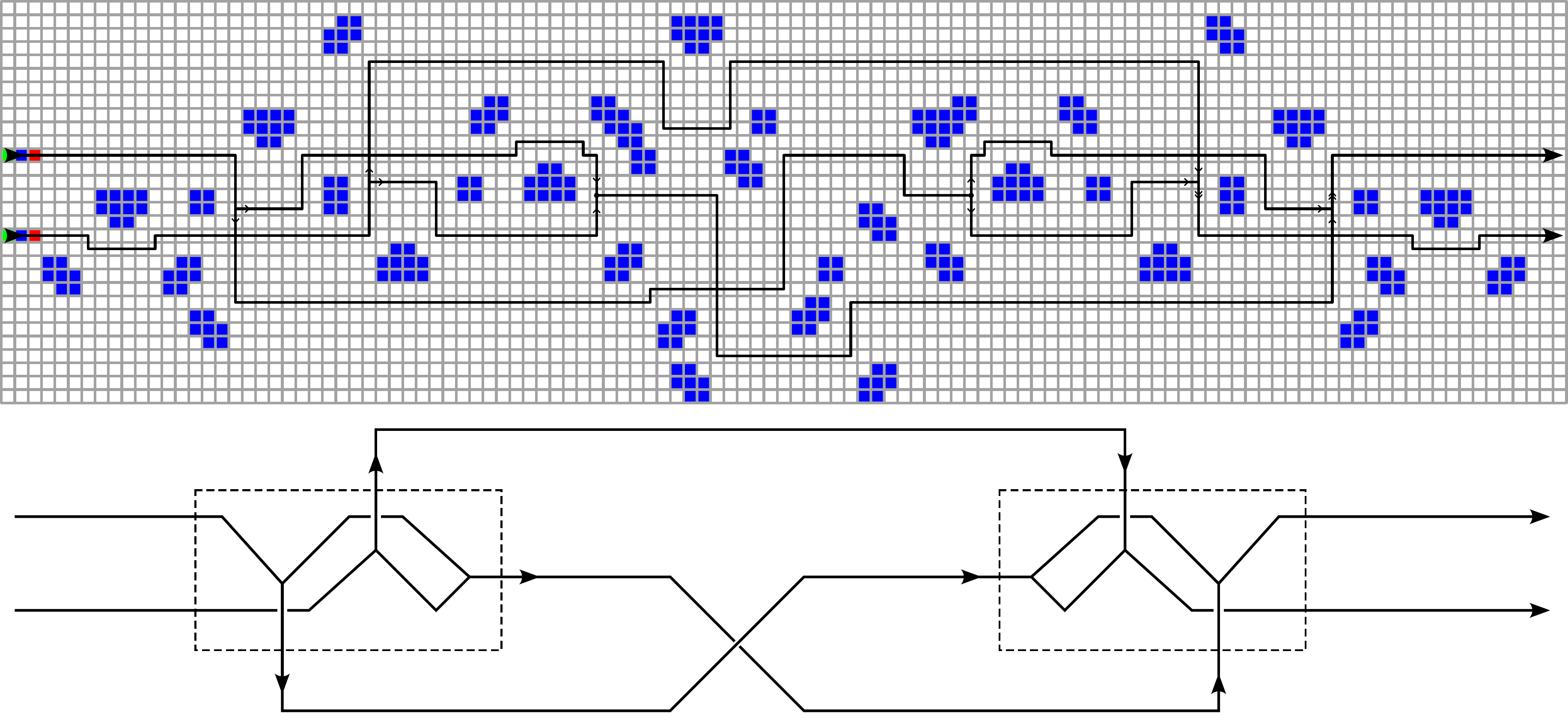}
\end{center}
\caption{Signals transformations in Controlled-NOT gate for such RCA as $\WL{1,2}$ }
\label{fig:c-not}
\end{figure}

Possible design is shown on \Fig{c-not}.
The simplified circuit diagram without delay elements shown at 
the bottom of \Fig{c-not} uses a gate depicted on \Fig{m-switch} (surrounded
by a dashed rectangle) together with inversion of this gate and exchanges paths 
of two signals between them.  

The Toffoli gate also can be designed for such RCA as $\WL{1,2}$, but
elaborated configuration is not shown here due to typographical and other reasons.
Anyway, the Toffoli gate can be composed from Fredkin and Controlled-NOT 
gates shown earlier on \Fig{fredkin} and \Fig{c-not}. 

\smallskip

Thus, considered family of CEOT RCA defined by property $n_c=0$
can be divided on two types with similar {\em regular stable} behaviour, 
but with some subtleties with {\em expansive} behaviour discussed further and
relevant for analysis of sensitivity and reliability. 
The first type may use fixed number of signals with different length
and for the second one the length is fixed, but signals 
can be split and merged.

Such a difference is mainly related with presence of pair $n_c=0$, 
$n_e=2$ in a sequence used for description of local rule.
As examples of first type further is used already
mentioned $\WL{1}$ together with $\WL{1,3}$
with analogue properties.
The second type together with $\WL{1,2}$ includes
$\WL{1,2,3}$ or $\WL{1-3}$ with similar properties
together with $\WL{1,2,3,4}$ or $\WL{1-4}$ denoted 
further simply as $\WL+$.

The RCA $\WL{1}$ was briefly mentioned first in Ref.~\cite{AV14},
$\WL{1,2}$, $\WL+$ and $\WL{1-3}$ were introduced and implemented 
in software developed by author with different names 
around the same time \cite{CART} together with yet another 
isotropic RCA. 

Last RCA formally is not in CEOT class, but it can be 
considered as minimal modification of $\WL{1,2}$ and denoted as $\WL{1,2o}$
due to additional requirement about two non-empty neighboring cells are 
being {\em opposite}. Similarly with $\WL{1,2}$ the $\WL{1,2o}$ also may be
used for gates with variable number of signals with minimal length (`ants'). 

It was already mentioned, that the second-order RCA can be considered as 
usual (first-order) RCA with four states. Using conversion into appropriate format 
for local rules with four states some numerical experiments discussed 
here was also performed on general software for CA simulation developed by 
A. Trevorrow, T. Rokicki, {\em et al} \cite{GoL}.

\section{Modelling of damage propagation}
\label{sec:damage}

\subsection{Damage propagation from single defect}
\label{sec:defect}
 
Quantification of damage propagation in CA can be useful for reliability analysis of
logic circuits based on such a model. Informal idea of {\em Lyapunov exponents} sometimes
used as a measure of such propagation \cite{WolCA} and few approaches to rigorous definition are developed \cite{BBR92,MS92,Ti00,BB10,BG18,Ko21} despite of certain 
difficulties and ambiguities related with application of such quantity to discrete systems.  
An intensity of defect accumulation and speed of the damage spread could be
mentioned as two alternative views on the Lyapunov exponent \cite{BG18}.

Let us note some subtleties with definition of exponential quantities in CA. 
For a single defect even for an extreme case of influence on all neighbors
after first step for $d$-dimensional CA there are $\epsilon=3^d$ affected cells 
in Moore neighborhood, but number of affected cells at $t$-th steps denoted further as
$\epsilon_t$ may grow only polynomially 
with time $\epsilon_t=(2t+1)^d$ and this amount is the maximally possible for any CA with 
local update rule. It provides natural constraint on exponential growth if 
to count defects as the total number of affected cells 
\begin{equation}
\label{expar}
\epsilon_t = O(e^{\lambda t}), \quad
\lambda = \lim_{t\to\infty}\frac{\ln \epsilon_t}{t}  \le 
\lim_{t\to\infty}\frac{\ln (2t+1)^d}{t}  = 
\lim_{t\to\infty}\frac{d}{t}\ln(2t+1)  =0\, .
\end{equation}

Such a problem could be addressed by consideration of separate replica of a system
for each defect for any time step \cite{BBR92} with formal use of so-called Boolean 
derivatives \cite{Vi90}, but such calculation without proper 
accounting of defects cancellation may produce overestimation of possible effects.

Let us consider 1D CA with two states and update rule
\begin{equation}
\label{rule90}
x_i^{(t+1)} = x_{i-1}^{(t)} \oplus x_{i+1}^{(t)},
\end{equation}
where $\oplus$ is addition modulo 2. According to enumeration in Ref.~\cite{W93} 
the CA corresponds to {\em rule 90} and provides important example of {\em additive} 
CA \cite{AlgCA}. Statistical properties of second-order RCA produced by such a rule  
denoted as 90R was discussed in Ref.~\cite{Ta87}. 

For initial (irreversible) rule 90 CA calculation with
Boolean derivatives produces value $\lambda = \ln 2$ for Lyapunov 
exponent  \cite{BBR92}. 
For configuration with single occupied cell the total number of non-empty 
cells after $n$ steps is described by sequence
A001316 in OEIS \cite{OEIS} and can be expressed as $2^{\wt{n}}$, where
$\wt{n}$ is number of units in binary expansion of $n$ (Hamming weight).

Despite of exponential notation the value $2^{\wt{n}}$ may not exceed $n+1$,
but for any even $n$ the value doubles on next step $n+1$ resembling
dynamics that could be expected for process with $\lambda = \ln 2$.
Damage front for such CA moves with unit speed in both directions
and could provide only minimal information for comparison with other 
CA \cite{BBR92}.

Local rule for similar 2D CA with two states and von Neumann neighborhood 
can be expressed as
\begin{equation}
\label{EdRep}
x_{i,j}^{(t+1)} = x_{i-1,j}^{(t)} \oplus x_{i+1,j}^{(t)}
 \oplus x_{i,j-1}^{(t)} \oplus x_{i,j+1}^{(t)}.
\end{equation}

This CA is notable as a simple example with self-replication of any pattern discovered
by E.~Fredkin about 1960 and recollected by M.~Gardner \cite[Ch. 21]{Gard}.
For a single occupied cell after $n$ steps there are $4^{\wt{n}}$ non-empty cells, see
sequence A102376 in OEIS. For particular values of $t$ the configurations
correspond to the square grids with different number of points and intervals between them. Due to \Eq{EdRep} the CA is linear with respect to addition modulo 2 and because
any pattern can be formally represented as a sum of few configurations
with single occupied cell, at appropriate time steps the pattern is replicated over grids
with sufficiently large intervals.

Using earlier conventions for CEOT rules such CA could be 
formally denoted as $\CET{0-4}{1,3}$, but the notation $\vNT{1,3}$ is more natural
for this CA with von Neumann neighborhood, 
because the local rule \Eq{EdRep} does not depend on number of cells with common corner.
The CA should be distinguished from similar replicator with Moore
neighborhood \cite{SlCA}, see sequence A160239 in OEIS.

\begin{figure}[htbp]
\begin{center}	
  \includegraphics[width=0.85\textwidth]{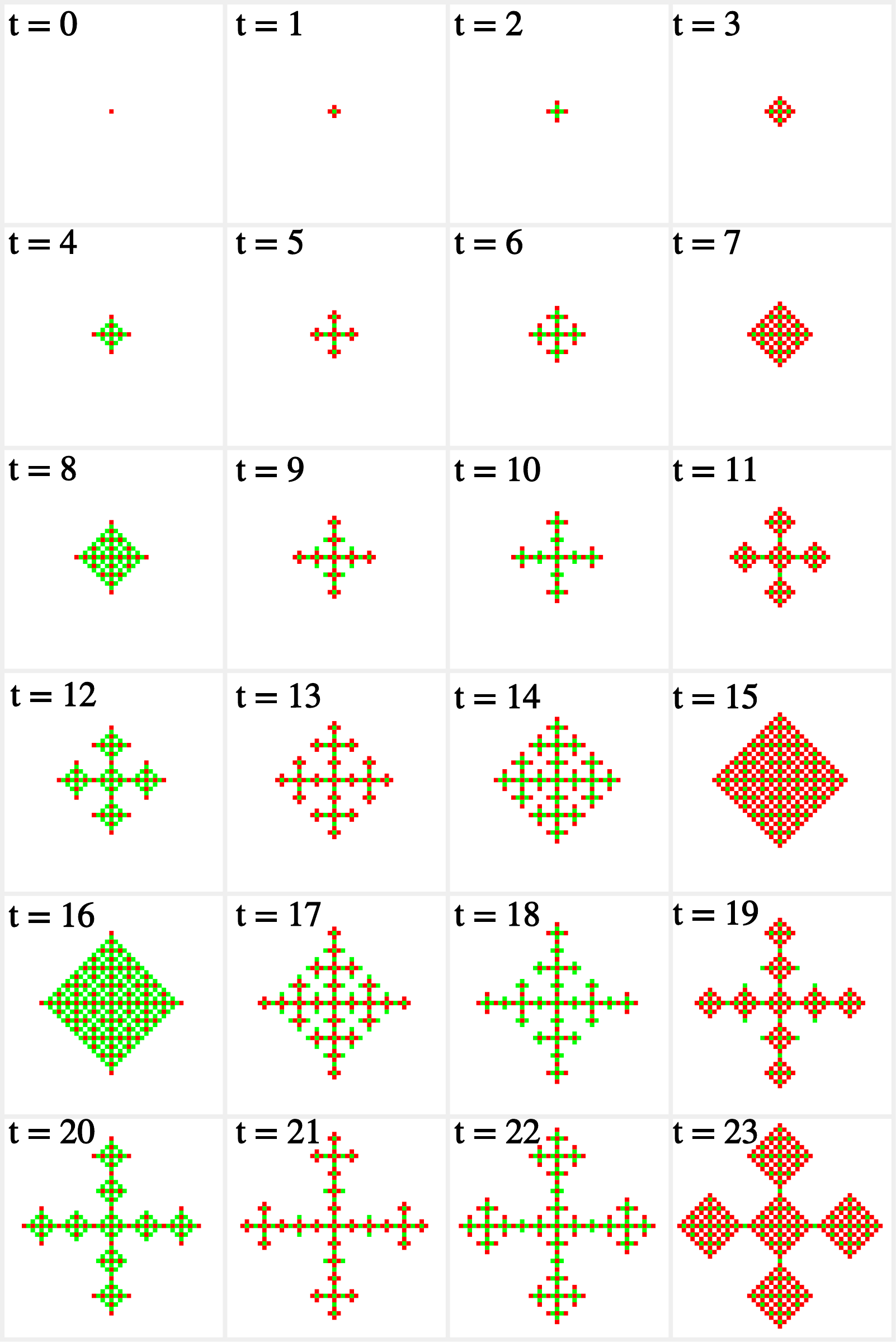}
\end{center}
\caption{Evolution of red cell in $\WL1$, $\WL{1,3}$ and $\REd$}
\label{fig:snakes}
\end{figure}

Second-order 2D RCA from $\vNT{1,3}$ can be denoted as $\REd$.
Number of cells with different states after $n$ steps for configuration
with a single cell with state 1 (red) is described by sequences
A244642 and A244643 in OEIS. 
The RCA $\REd$ is not replicator, but it may produce interesting
fractal-like patterns.
For configuration with single red cell
evolution is also coincides with considered earlier RCA $\WL1$ \cite{AV14} and
for $\WL{1,3}$ such coincidence also can be shown in a similar way due to 
some similarity of local rules with $\WL1$ and $\REd$, see \Fig{snakes}.

The numbers $N_1$ and $N_2$ of cells with states 1 and 2 after $n$ steps for configuration
with single red cell satisfy recursive formulas \cite{AV14}
\begin{equation}
\label{NREd} 
\begin{split}
 N_1(2n+1) &= 4 N_1(n),\quad  N_1(2n+2)  = N_1(n) + N_1(n+1), \\
 N_2(2n) &= 4N_2(n),\quad  N_2(2n+1) = N_2(n) + N_2(n+1).
\end{split} 
\end{equation}
The first part of each equation would correspond to asymptotically
quadratic growth $O(n^2)$, but the second part to only linear one $O(n)$. Such a 
variance in formal estimation of dimensions seems to be in some agreement with
the shape of configurations \Fig{snakes} resembling some geometrical sequences used
for construction of fractals \cite{Frac}.

Such behaviour justifies consideration of models with polynomial growth together 
with an analogy of formal or fractal dimension \cite{Frac} already investigated 
earlier in context of CA \cite[Ch. 4]{Ila} and denoted further as $\Dfr$ that 
is not vanishing unlike $\lambda$ \Eq{expar}
\begin{equation}
\label{polpar}
\epsilon_t = O(t^\Dfr), \quad \Dfr(t) = \frac{\ln \epsilon_t}{\ln t} > 0. 
\end{equation}
The subtleties of transition from definition of fractal dimension relevant to continuous
case to such discrete systems as CA can be found in Ref~\cite[Ch.~4]{Ila}, but here 
{\em formal dimension} $\Dfr$ is treated rather as convenient quantity defined by \Eq{polpar}.

Let us also define slightly different quantity
\begin{equation}
\label{Ddpar}
\Ddfr_t = \frac{\ln(\epsilon_t)}{\ln(\sqrt{2} t)} 
= \frac{\ln(\epsilon_t)}{\ln(t)+\ln(2)/2}. 
\end{equation}
Compared to the earlier definition \Eq{polpar} $\Ddfr_t$ includes 
extra multiplier $\sqrt{2}$ for better adapting for diamond-like 
shape representing maximal possible damage spread for such kind of RCA.
The number of cells for maximal possible damage is close to area
of such diamond $N \appreq S = 2t^2$ and so formal normalization for $\Ddfr=1$ should 
fit with $\sqrt S = \sqrt 2 t$, but term $\ln(2)/2 \appreq 0.35$ 
in \Eq{Ddpar} looks relevant only for relatively small $\ln(t)$.

These parameters meet obvious inequality $\Ddfr_t \le \Dfr_t$, but if the
limits for $t\to\infty$ exist, they are equal and denoted further as
\begin{equation}
\label{Dinf}
 \Dfr_\infty \equiv \lim_{t \to \infty}\Dfr_t = \lim_{t \to \infty}\Ddfr_t.
\end{equation}

For $\WL1$, $\WL{1,3}$ and $\REd$ for configuration
with a single red cell the limit $\Dfr_\infty$ \Eq{Dinf} does not exist,
because for $\epsilon_t = N_1(t)+N_2(t)$ 
due to \Eq{NREd} $\Ddfr$ oscillates between $\Ddfr_{\max}=2$ and 
$1 < \Ddfr_{\min} < 2$, $\lim_{t\to\infty}\Ddfr_{\min} =1$.
The Table~\ref{Dmin} illustrates rather slow decrease of 
$\Ddfr_{\min}$ defined as minimal value of $\Ddfr_t$
for given $t$ (starting from initial value $t=1$ instead 
of $t=0$ used here in some illustrations). 

\begin{table}[htbp]
\[
 \begin{array}{|r|r|l|}
 \hline 
 t & \epsilon_t & \Ddfr_{\min} \\ \hline
       2 &          5 &  1.54795206 \\   
       3 &          9 &  1.52037507 \\
      22 &        181 &  1.51223900  \\
      38 &        405 &  1.50693934   \\
      42 &        469 &  1.50593450   \\
      43 &        441 &  1.48232237   \\
      86 &       1165 &  1.47065043   \\
     171 &       2929 &  1.45445908   \\
     342 &       7589 &  1.44538094   \\
     683 &      19305 &  1.43576621   \\
    1366 &      49661 &  1.42911275   \\
    2731 &     126881 &  1.42281264   \\
    5462 &     325525 &  1.41789438   \\
   10923 &     833049 &  1.41343346   \\
   21846 &    2135149 &  1.40970770   \\
   43691 &    5467345 &  1.40636763   \\
   87382 &   14007941 &  1.40346873   \\
  174763 &   35877321 &  1.40086589   \\
  349526 &   91909085 &  1.39855394   \\
  699051 &  235418369 &  1.39646489   \\
 1398102 &  603054709 &  1.39458094   \\
 2796203 & 1544728185 &  1.39286577   \\
 5592406 & 3956947021 &  1.39130212   \\
11184811 & 10135859761 &  1.38986817   \\
22369622 & 25963647845 &  1.38854995   \\ 
\hline
\end{array}
\]
\caption{$\Ddfr_{\min}$ for $\REd$ configurations starting from single red cell}
\label{Dmin} 
\end{table}

\begin{figure}[htb]
\begin{center}	
  \includegraphics[width=0.85\textwidth]{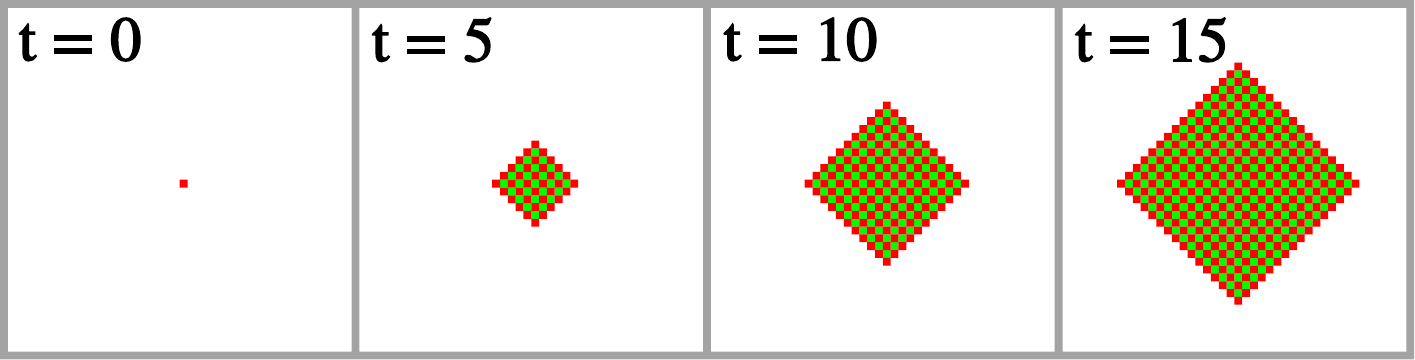}
\end{center}
\caption{Evolution of red cell in $\WL{1,2}$ and $\WL{1-3}$}
\label{fig:ants}
\end{figure}

\begin{figure}[htb]
\begin{center}	
  \includegraphics[width=0.85\textwidth]{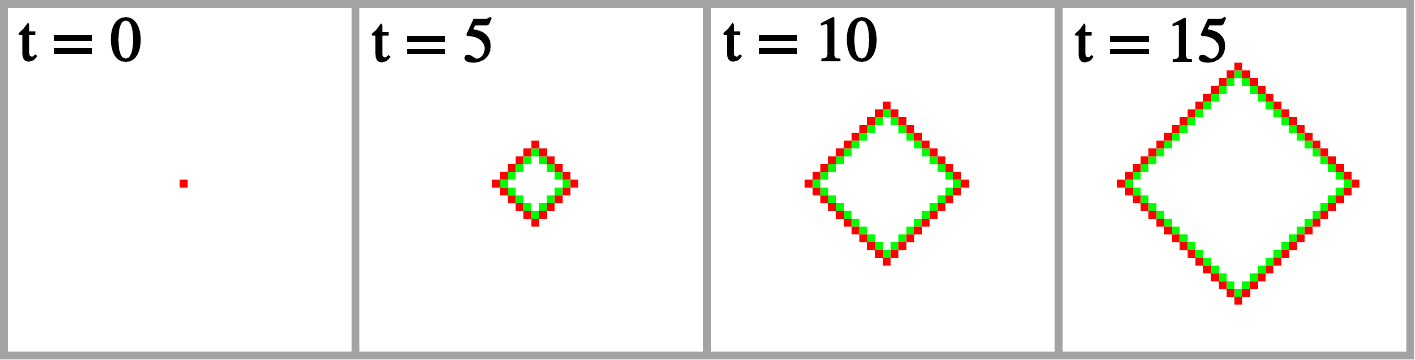}
\end{center}
\caption{Evolution of red cell in $\WL+$}
\label{fig:ants2}
\end{figure}

The evolution of patterns from a single red cell for other RCA 
can be quite simple
with $\Dfr_\infty = 2$ for $\WL{1,2}$ and $\WL{1-3}$ depicted on \Fig{ants} and
$\Dfr_\infty = 1$ for $\WL+$ depicted on \Fig{ants2}.  

\begin{figure}[htbp]
\begin{center}	
  \includegraphics[width=0.85\textwidth]{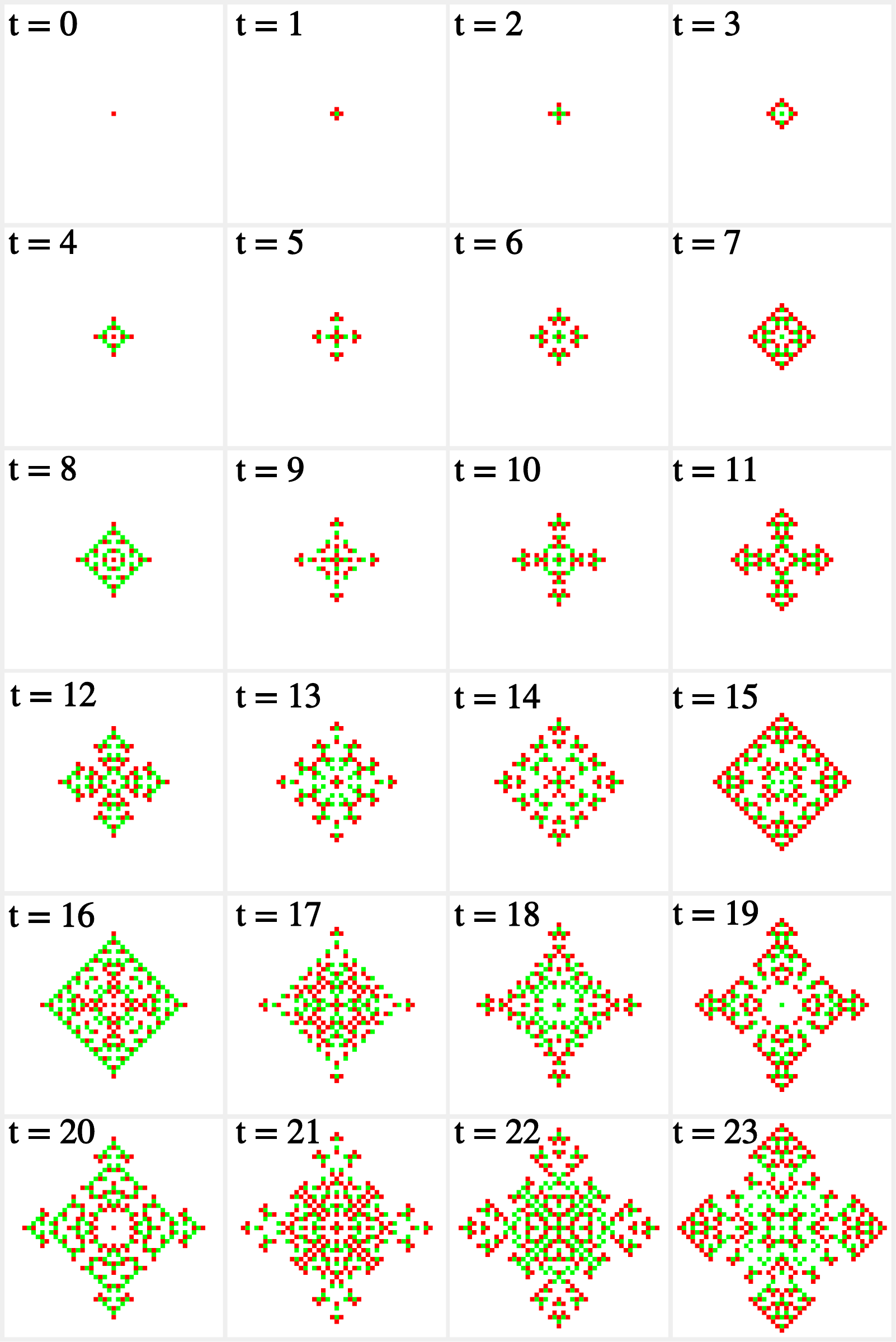}
\end{center}
\caption{Evolution of red cell in $\WL{1,2o}$}
\label{fig:ants4}
\end{figure}

Evolution of isotropic RCA $\WL{1,2o}$ from single red cell is
shown on \Fig{ants4}.
Analytical expression for number of cells is not
known to author, but numerical estimation of $\Ddfr_t$ 
shows slow growth 
with minimal variations and upper limit by some value $\Ddfr_{\max} \le 2$. 
For example, $\Ddfr_{\max}$ for $\WL{1,2o}$ is increasing 
from $\Ddfr \appreq 1.55$ for $t=2$ to $\Ddfr \appreq 1.88$ for $t=1000$.

However, behaviour becomes less trivial for more complex configurations. For example
static $2 \times 2$ blue block added to initial configuration with single
red cell does not change values $\Dfr_\infty=2$ for $\WL{1,2}$ and 
$\Dfr_\infty=1$ for $\WL+$, but for $\WL{1-3}$ growing rectangular 
area of empty cells appears inside of initial diamond patters resulting 
$\Dfr_\infty=1$.

Behaviour looks less predictable if to add static configurations with more complex shapes.
In such a case $\Ddfr$ for $\WL+$ can increase from $\Dfr_\infty=1$ to 
$\Dfr_\infty=2$ due to many
interfering damage fronts and for $\WL{1-3}$ earlier mentioned decrease from 
$\Dfr_\infty=2$ to $\Dfr_\infty=1$ also
can be suppressed due to similar effects. In both cases limiting value may depend
in some non-trivial way from a shape of static configuration and initial position 
of red cell.

\subsection{Damage from improper signals interaction}
\label{sec:interact}

Despite of possible convenience for consideration of damage propagation due to certain 
simplification, the point-like defect discussed in \Sec{defect} could not appear during normal 
evolution of RCA. Let us consider some kind of errors due to incorrect interaction between 
signals such as wrong relative delays or too short intervals between consequent groups of signals.

For example, all signals should enter in a gate at the same time. A few kinds 
of errors could be expected due to delay between signals. The delay of output signals
could be considered as a minimal possible problem. The simple example is the swap
gate \Fig{swap}. However, even for swap gates there is critical range of
delays producing more serious problems. For `wide' swap gate such a critical
range for delay of second (lower) signal is $3 - 11$ time steps  and for 
`narrow' swap the range is even bigger ($3 - 15$), because signal trajectories 
intersect repeatedly.

The similar errors can appear if an interval between two consequent groups of
signals is too small. For example, the interval should be at least 12 for `wide' 
swap and 16 for `narrow' swap.

Errors also may depend on the particular value of delay and used RCA.
Let us consider the interaction of two signals approaching in orthogonal
directions to the point of intersection of trajectories. The difference
of distances between signals and this point can be denoted as $\Delta$.
For minimal length of signals \mbox{($k=3$)} some interaction occurs
for $-4 \le \Delta \le 4$. 
It explains value of critical range for `wide' swap, because the first 
signal reach intersection point with additional delay 7 steps in
comparison with the second one.

Collisions of signals with different delays are shown for discussed
in this work RCA on \Figs{snakes-delays}{ --}{ant4-delays}
in \ref{AppA}.

\smallskip

The properties of simple collisions may only provide some guesses about behaviour
in more complex elements and computer simulations are necessary to clarify details. 
For already mentioned `wide' swap the delay 7 time steps between
two signals corresponds to $\Delta=0$ and produces reflection with reverse
motion of signals for $\WL1$, $\WL{1,3}$  and $\WL{1,2o}$, but for
$\WL{1,2}$, $\WL{1-3}$ and $\WL+$ the damage is more serious resembling
behaviour for boundary values 3 and 11 of critical range.

For `snakes' $\WL1$, $\WL{1,3}$ the delay values $(4,5,6,8,9,10)$ 
which are not equal to 
ends $(3,11)$ or middle $(7)$ of critical range correspond to generation of 
expanding signal after collision, but interaction with static elements produces 
a signal with minimal length (`ant') and a `snake' with variable length and 
directions depending on particular values of delay.

For all `ants' $\WL{1,2}$, $\WL{1-3}$, $\WL+$ and $\WL{1,2o}$ collision of signals 
with delays $(4,5,6,8,9,10)$ produces three `ants', but due to interaction after split 
resulting configuration includes three signals only for values $(4,6,8,9,10)$ with the 
particular case of delay 5 resulting omnidirectional damage front with complex shape 
due to interaction of two signals.

Last case also demonstrates ambiguities with characterization using formal dimensions, 
because even improper signal routing formally characterizing for
unconstrained motion by $\Dfr_\infty \appreq 0$ 
may produce further a damage front with $1 \le \Dfr_\infty \le 2$ due to interaction 
with static elements and other signals.

\smallskip

Intersection of signals in swap gates or other kinds of routing can be considered as 
particular problem of 2D design and resolved in 3D RCA models with similar properties. 
However for any non-trivial gate discussed above interaction of signals is necessary 
process. An elementary interaction relevant already for simple switch gate
\Fig{switch} can be considered to investigate errors due to wrong delays.
Such interactions are depicted on \Figs{snakes-bl-delays}{ --}{ant4-bl-delays}
in \ref{AppB}.

\subsection{Consequences of damage distribution}
\label{sec:eff}

Description of damage distribution is also important because improper
interaction of signals in some gate may produce undesirable effects
on whole circuit. For $\Dfr_\infty=2$ result of such error should
disrupt the proper work of other elements then it reaches them. 

For $\Dfr_\infty = 1$ omnidirectional damage front (see \Fig{ants2})
also may produce the similar disruption of functionality. An alternative
example of $\Dfr_\infty = 1$ with anisotropic distribution can be illustrated
by result of collision RCA $\WL1$ or $\WL{1,3}$ (`snakes') with 
static block. In such a case the signal starts expansion in two
opposite direction. If only one of two `heads' of expanding
line reaches some obstacle further evolution correspond to motion
of a signal in unconstrained direction.

Collision of such signal with some static element results to
contraction of the signal to a size of single cell with further
expansion in the perpendicular direction. Thus, interaction
of {\em single} signal with {\em static} elements for $\WL1$ or $\WL{1,3}$ may
finally either produce expanding line with $\Dfr_\infty = 1$
or unconstrained signal motion formally resulting to $\Dfr_\infty = 0$.

However, even for $\WL1$ or $\WL{1,3}$ already interaction of {\em two} 
signals may produce omnidirectional spreading of damage 
with $\Dfr_\infty = 1$ or $\Dfr_\infty = 2$  resulting disruption of circuit 
functionality (see \Figs{snakes-delays},{snakes3-delays} in \ref{AppA}
or \Figs{snakes-bl-delays},{snakes3-bl-delays} in \ref{AppB} for $|\Delta|=4$).

For other RCA such as $\WL{1,2}$,  $\WL+$, $\WL{1-3}$ and $\WL{1,2o}$ (`ants') 
even for one signal
collision with static block leads to the appearance of two signals moving 
in opposite directions. Unconstrained motion of such signals formally would 
result to $\Dfr_\infty = 0$, but consequent collisions with static elements 
produce a fast increase in the number of signals. The interaction between such 
signals again may produce omnidirectional damage spread (see 
\Figs{ant-delays}{~--}{ant4-delays} in \ref{AppA} and \Figs{ant-bl-delays}{ --}{ant4-bl-delays}
in \ref{AppB}) generating errors in the gates affected by its propagation.

Parts of circuits can be isolated from an external damage using static elements,
but such protection makes impossible an interaction between them. Signal input and output
also becomes problematic with such isolation. Numerical experiments
also demonstrate non-effectiveness of static elements or even movable `covers' 
designed to protect against destructive noise between consequent signals.

\smallskip

Thus, for RCA considered in this work reversible circuits are very 
sensitive to errors. Delays of signals or too short interval between groups 
of signals even in one gate may result appearance of omnidirectional damage 
that prevents further correct operation of whole reversible logic 
circuits.

For example, already discussed minimum intervals between consequent
signals for `wide' and `narrow' swap gates on \Fig{swap} should be
12 and 16 time steps respectively. For Fredkin gate on \Fig{fredkin}
minimum interval depends on the composition of the signal (representing 
combinations of zeros and units) in two consequent groups.
In the worst case such interval should not be less than 22 time steps.
Controlled-NOT gate on \Fig{c-not} can be used for RCA  
$\WL{1,2}$, $\WL+$, $\WL{1-3}$ and $\WL{1,2o}$ with a relatively
short minimum possible interval of 10 time steps between consequent 
groups of signals.

\section{Discussion and conclusions}
\label{sec:fin}

Let us recollect properties of CA and RCA discussed in this work.
\begin{center}
\begin{tabular}{|l|c|c|c|c|c|c|c|}\hline
Properties $\smash{\!\!\diagdown\!\!}$ CA
   & $\!\CET01\!$ & $\!\CET0{1,3}\!$ &$\!\CET0{1,2}\!$ &$\!\CET0+\!$&$\!\CET0{1-3}\!$
   & $\!\CET0{1,2o}\!$&$\!\vNT{1,3}\!$ \\ \hline
Isotropic   
   & yes      & yes          & yes         & yes    & yes
   & yes      & yes \\
CET/ET
   & CET      & CET          & CET         & CET    & CET
   & no       & ET \\
Outer   
   & yes      & yes          & yes         & yes    & yes
   & yes      & yes \\
Linear  
  & no        & no           & no         & no    & no
  & no        & yes \\
Replicator 
  & no        & no           & no         & no    & no
  & no      & yes \\  \hline
\multicolumn{7}{|c|}{Logic gates in second-order RCA} &
\multicolumn{1}{c|}{}\\ \cline{1-7}
Swap   
   & yes      & yes          & yes         & yes    & yes
   & yes      &  \\
Fredkin   
   & yes      & yes          & yes         & yes    & yes
   & yes      &  \\
C-NOT
   & no      & no          & yes         & yes    & yes
   & yes     &  \\   
\hline      
\end{tabular}
\end{center}

Property `outer' emphasizes already explained in \Sec{class} class of 
`inner-independent' CA. The shorter notation ET (edge-totalistic) is used 
instead of CET (corner-edge totalistic) for $\vNT{1,3}$ formally
defined for von Neumann neighborhood excluding cells with common
corners unlike CA with Moore neighborhood.

Despite of impossibility to use  $\REd$ for construction of 
reversible circuits this CA is included in the comparison due to known remarkable 
properties such as linearity and self-replication of any pattern in $\vNT{1,3}$
\cite{Gard}. All CA used here for construction of second-order RCA could be 
informally treated as some non-linear 
modifications of the Fredkin replicator $\vNT{1,3}$.

The analytical formulas \Eq{NREd} obtained due to linearity of $\REd$ 
also can be extended to  $\WL1$ \cite{AV14} (and $\WL{1,3}$) to analyse simple
example of damage distribution. Oscillating behaviour of {\em formal dimension}
$\Dfr$ understandable from such expressions provides some hints about possible issues 
with instability in damage distribution for considered family of RCA partially confirmed 
by numerical experiments.

Due to such instability the RCA family has high sensitivity to
errors, because even local defects or improper functionality 
of single gate finally leads to failure of the entire circuit.
However, such behaviour also provides simple error indication.
A fault almost inevitably produces quickly recorded growth in 
total number of non-empty cells signaling a failure and 
preventing the completion of calculations with an incorrect result.

On the other hand, the serious damage from even a single error, along 
with the difficulty of isolation other elements from its influence, 
lead to problems with longer reliable work of reversible circuits with
many gates partially consistent with conclusion drawn 
from altentative model with probabilistic RCA \cite{RCAnoise}. 

The problem with sensitivity to errors may be partially
related with constructions of considered RCA. They are 
adapted for simpler design of logic gates, but
not for reliability. The local rules of RCA 
considered in this work provide a natural way to create small 
mobile configurations that easily change direction of movement.
At the same time, close similarity with RCA $\REd$ derived
from Fredkin replicator could result extra instability due
to variance of formal dimension $\Dfr$ already discussed 
earlier in \Sec{damage}.

\smallskip

Only few RCA from rather big families such as CEOT are discussed here. 
However for construction of reversible logic circuits even one of them 
such as $\WL{1,2}$ would be enough to demonstrate examples with both 
conserved and variable number of signals. Other RCA discussed here
are useful for analysis of reliability and damage distribution.

For all RCA presented in this work computational experiments demonstrate
too serious problems for whole circuits even from single errors.
Possibility to reduce such effects could be considered as a
reasonable research direction.

Perhaps, the choice of local 
rules aimed at greater stability of configurations could 
reduce sensitivity to errors. For example, even small difference
between $\WL{1,2}$ and $\WL{1,2o}$ hinders the damage spread 
after side impact with zero delay, {\em cf}\/ \Fig{ant-delays} with 
\Fig{ant4-delays} for $\Delta=0$ in \ref{AppA}.

The $\WL{1,2o}$ belongs to even wider family, than CEOT. It is
second-order RCA derived from isotropic inner-independent CA. 
The total number of isotropic CA mentioned earlier is 
$2^{102} \appreq 5 \times 10^{30}$ and number of inner-independent 
isotropic CA is $2^{51} \appreq 2 \times 10^{15}$.
Thus, brute force method is hardly appropriate to search
for more reliable RCA.

There is simple measure of complexity for such local rules from
CET and CEOT family based on counting number $N_p$ of mentioned
in \Sec{class} pairs $p=(n_c,n_e)$ with $f(p)=1$ producing change
of central cell. For example, $N_p=1$ for $\CET01$ or $\WL1$ and 
$N_p=2$ for $\WL{1,2}$. For isotropic rules similar number $N_I$ 
of different configurations up to rotations and reflections should 
be used instead.

For example, pair $(n_c=0,n_e=2)$ includes two different
isotropic configurations with two neighboring cells are either
opposite or adjacent. The first case (with opposite 
cells) is used in definition of $\WL{1,2o}$. Thus, $N_I=1$ for $\WL1$,
$N_I=2$ for $\WL{1,2o}$ and $N_I=3$ for $\WL{1,2}$.

For such complexity measures the universal computation can be 
obtained already for `simplest' RCA $\WL1$, $\WL{1,2o}$ with 
$N_I=1, 2$. More complex 
local rules may be necessary for investigation of stability and 
damage distribution, but each new pair or configuration included
in the rule may produce destruction of initially stable static 
elements used for signal routing or undesirable interactions
of signals with such elements. Such problems with more
complex local rules could be partially avoided by using special 
passive `mirrors' or active `wires' for signal routing.

\smallskip

Similarly with BBM (billiard ball model of computations) RCA discussed in this 
work do not use any `wires' for motion between gates. Analogue models with 
`wires' could be designed using generalization of second-order RCA \cite{AV13} 
implemented in software mentioned earlier \cite{CART}. Such RCA 
have five states with additional one for construction of `wires'. That could 
reduce sensitivity to external noise, because all cells around wires
are passive and may not participate in damage distribution. 
Reliability analysis for such RCA 
with damage distribution only along wires should be 
discussed elsewhere. 

The existence of alternative models with `wires' also justifies earlier informal 
name `wireless ants' for some RCA discussed here and abbreviations 
for local rules in software \cite{CART} such as WlAnt, WlAnt-2, WlAnt-3 and WlAnt-4 
for $\WL{1,2}$, $\WL+$, $\WL{1-3}$ and $\WL{1,2o}$ respectively.
The local rules for $\WL1$ (WlSnake) and $\WL{1,3}$ (WlSnake-3) are  
implemented in the same package. All 
rules are also converted to so-called `tree' format used for
CA simulations in more common software \cite{GoL}.

\appendix
\setcounter{figure}{0}

\section{Modelling of signals collision}
\label{AppA}

\begin{figure}[htbp]
\begin{center}	
  ~\includegraphics[width=0.855\textwidth]{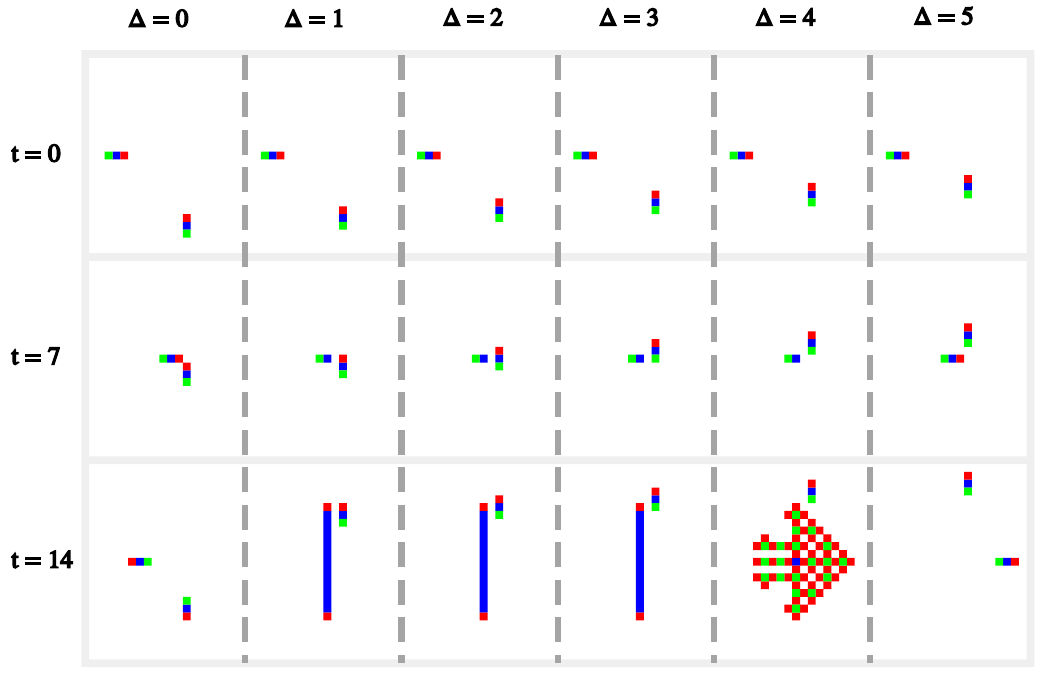}
\end{center}
\caption{Collisions with different delays in $\WL1$}
\label{fig:snakes-delays}
\end{figure}

\begin{figure}[htbp]
\begin{center}	
  \includegraphics[width=0.85\textwidth]{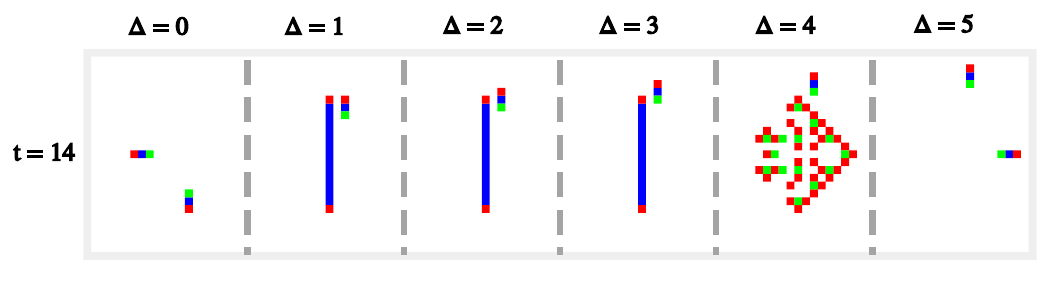}
\end{center}
\caption{Collisions with different delays in $\WL{1,3}$}
\label{fig:snakes3-delays}
\end{figure}

\begin{figure}[htbp]
\begin{center}	
  \includegraphics[width=0.85\textwidth]{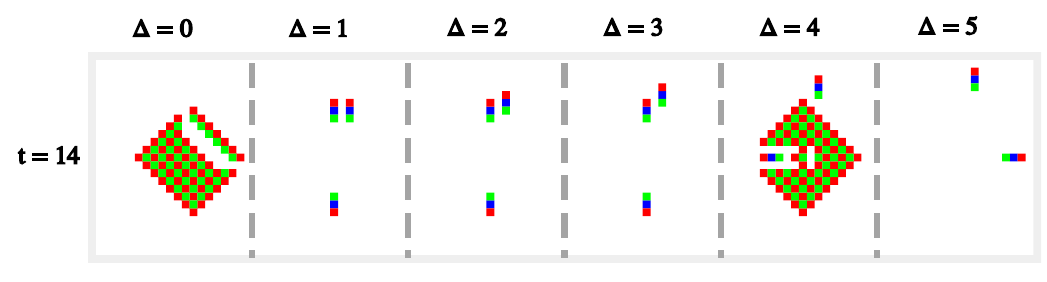}
\end{center}
\caption{Collisions with different delays in $\WL{1,2}$}
\label{fig:ant-delays}
\end{figure}

\begin{figure}[htbp]
\begin{center}	
  \includegraphics[width=0.85\textwidth]{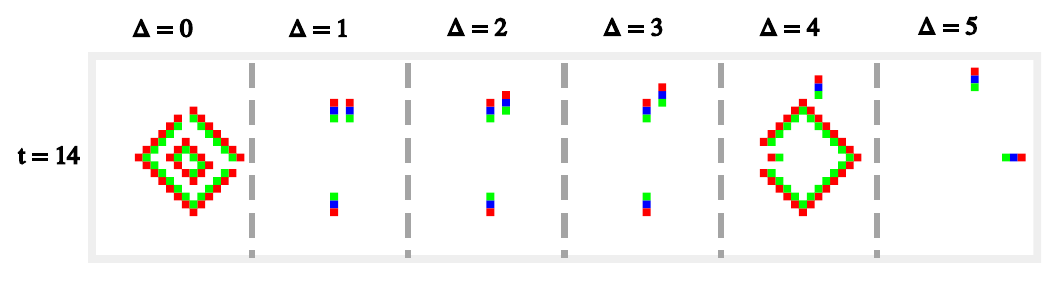}
\end{center}
\caption{Collisions with different delays in $\WL+$ and $\WL{1-3}$}
\label{fig:ant23-delays}
\end{figure}

\begin{figure}[htbp]
\begin{center}	
  \includegraphics[width=0.85\textwidth]{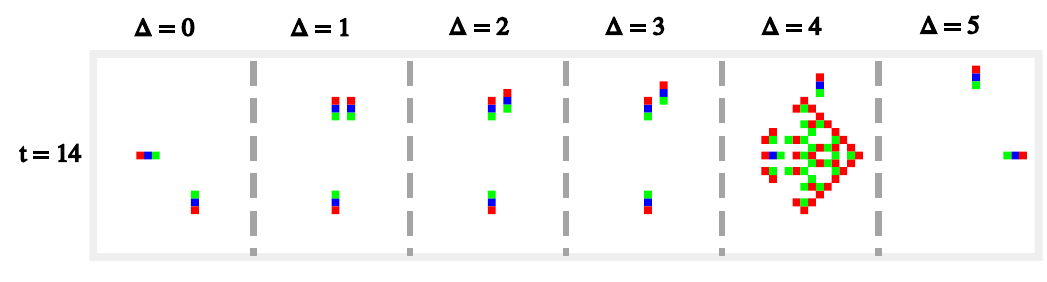}
\end{center}
\caption{Collisions with different delays in $\WL{1,2o}$}
\label{fig:ant4-delays}
\end{figure}

Collisions of signals with different delays $\Delta$ are shown for discussed
in this work RCA on Figures~\ref{fig:snakes-delays} -- \ref{fig:ant4-delays}
for $t=0$, $t=7$ and $t=14$.
The configurations for $t=0$ and $t=7$ are equal for all RCA
and shown only on Figure~\ref{fig:snakes-delays}.
Due to symmetry of picture with respect to reflections about the diagonal 
only six values $0 \le |\Delta| \le 5$ are shown for each case. The value
$|\Delta| = 5$ is the minimal one without any interaction.

For RCA $\WL1$ and $\WL{1,3}$ of the first type sometime referred earlier 
as `snakes' depicted on \Fig{snakes-delays} and \Fig{snakes3-delays} 
the value $\Delta=0$ corresponds to reflection of signals followed
by motion in opposite direction. The values $1 \le |\Delta| \le 3$ produce
a signal expanding in two opposite directions together with initial one.
The value $|\Delta| = 4$ seems produce more critical damage
with front expanding in all possible directions. 

The same value $|\Delta| = 4$ also looks critical for second type
of RCA $\WL{1,2}$, $\WL{1-3}$, $\WL+$ and $\WL{1,2o}$ sometimes referred earlier as 
`ants' depicted on \Fig{ant-delays} -- \Fig{ant4-delays}, but only for $\WL{1,2o}$
value $\Delta = 0$ produces simple reflection and for other
four RCA of such type $\Delta = 0$ causes more critical omnidirectional damage spread.
The values $1 \le |\Delta| \le 3$ for all four RCA of such type due to split of one signal 
after collision produce three signals moving apart with apparently 
smaller damage.

Both $\WL{1-3}$ and $\WL+$ produce the same damage patterns \Fig{ant23-delays} with 
$\Dfr_\infty = 2$ for $\Delta = 0$ or $\Dfr_\infty = 1$ for $|\Delta| = 4$, 
despite of different behaviour 
for initial configuration with single red cell discussed earlier,
see \Fig{ants} and \Fig{ants2}.

\section{Modelling of signals interaction}
\label{AppB}
\setcounter{figure}{0}

\begin{figure}[htbp]
\begin{center}	
  ~\includegraphics[width=0.855\textwidth]{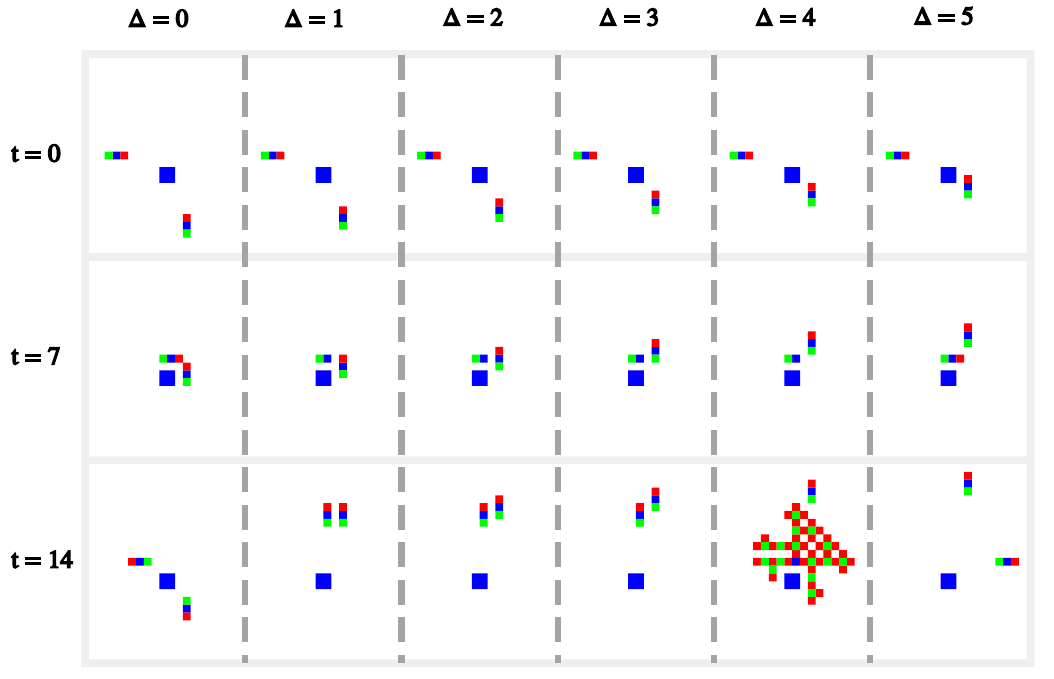}
\end{center}
\caption{Interactions with different delays in $\WL1$}
\label{fig:snakes-bl-delays}
\end{figure}

\begin{figure}[htbp]
\begin{center}	
  \includegraphics[width=0.85\textwidth]{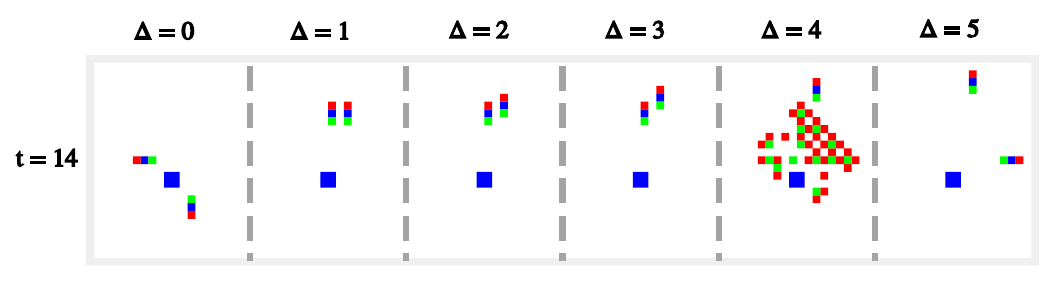}
\end{center}
\caption{Interactions with different delays in $\WL{1,3}$}
\label{fig:snakes3-bl-delays}
\end{figure}

\begin{figure}[htbp]
\begin{center}	
  \includegraphics[width=0.85\textwidth]{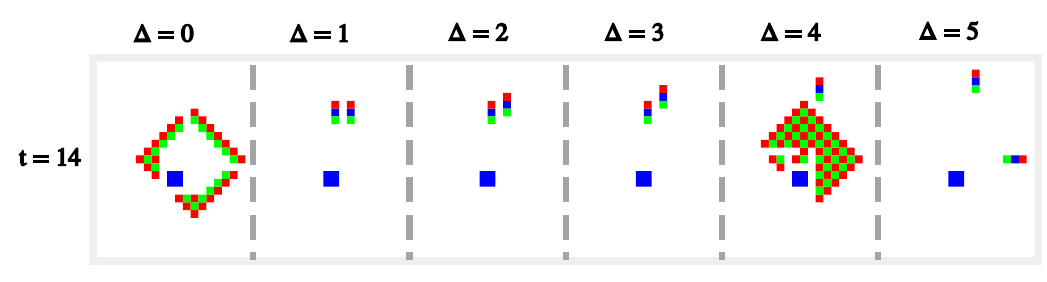}
\end{center}
\caption{Interactions with different delays in $\WL{1,2}$}
\label{fig:ant-bl-delays}
\end{figure}

\begin{figure}[htbp]
\begin{center}	
  \includegraphics[width=0.85\textwidth]{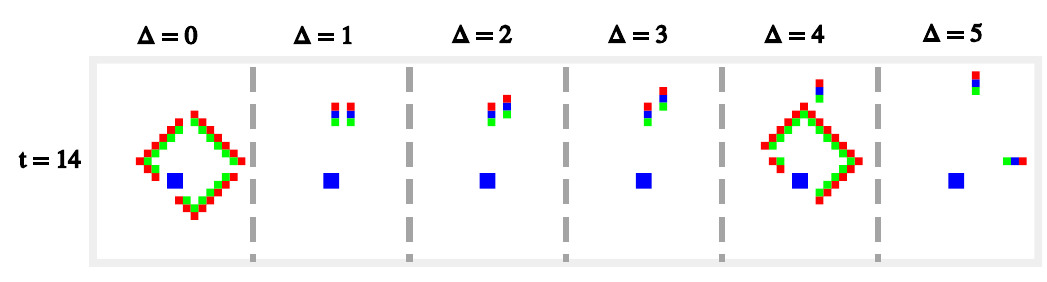}
\end{center}
\caption{Interactions with different delays in $\WL+$ and $\WL{1-3}$}
\label{fig:ant23-bl-delays}
\end{figure}

\begin{figure}[htbp]
\begin{center}	
  \includegraphics[width=0.85\textwidth]{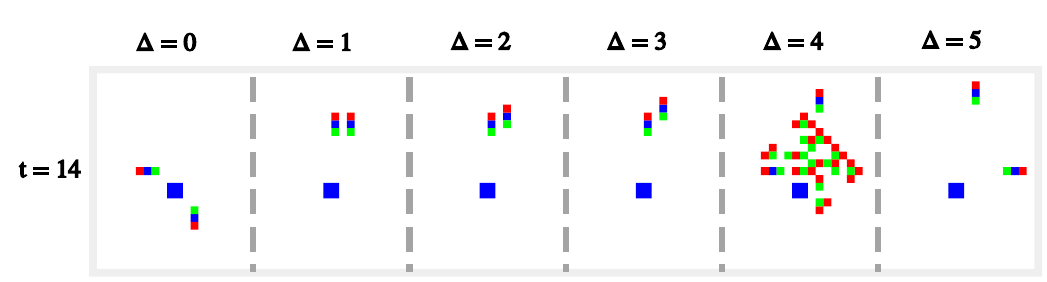}
\end{center}
\caption{Interactions with different delays in $\WL{1,2o}$}
\label{fig:ant4-bl-delays}
\end{figure}

An elementary interaction of signals is depicted on 
Figures~\ref{fig:snakes-bl-delays} -- \ref{fig:ant4-bl-delays}.
The configurations for $t=0$ and $t=7$ again are equal for all RCA
and shown only on Figure~\ref{fig:snakes-bl-delays}.
The main difference in comparison with Figures~\ref{fig:snakes-delays} -- \ref{fig:ant4-delays}
is $2 \times 2$ static blue block (cells with state 3) for proper signal motion after collision. 
The interaction between signals again corresponds to \mbox{$-4 \le \Delta \le 4$} and six values $0 \le |\Delta| \le 5$ are enough 
for each case due to symmetry 
with respect to reflections about the diagonal. 

For $1 \le |\Delta| \le 3$  after interaction two signals move along parallel paths
with relative delay $|\Delta| - 1$.
For elements similar with switch gate \Fig{switch} proper value is $|\Delta|=1$ 
taking into account necessity for repeated interaction.
The value $|\Delta|=4$ produces omnidirectional damage front
for all RCA discussed in the paper and $\Delta=0$ is related
with similar problem for $\WL{1,2}$, $\WL{1-3}$ and $\WL+$.
For $\WL1$, $\WL{1,3}$ and $\WL{1,2o}$ the same value $|\Delta|=0$
causes reflection of both signals after interaction.

Both $\WL{1-3}$ and $\WL+$ produce the same damage fronts with 
\mbox{$\Dfr_\infty = 1$}, see \Fig{ant23-bl-delays}.
Despite of some resemblance for $\Delta=0$ of initial time steps 
shown on \Fig{ant-bl-delays}, $\Dfr_\infty = 2$ for $\WL{1,2}$ not 
only for $|\Delta|=4$, but also for $\Delta=0$ with the shape of damage
described as a diamond-like area with rectangular empty space due to initial
impact of $2 \times 2$ static blue block.

\section*{Acknowledgements}
I would like to thank the anonymous reviewers for
their constructive comments and suggestions.

\end{document}